%% file: maintext.tex
\documentclass[
reprint,
showpacs,
superscriptaddress,
prl,
amsmath,amssymb,
aps,
]{revtex4-1}

\usepackage{graphicx}
\usepackage{dcolumn}
\usepackage{bm}
\usepackage{hyperref}
\usepackage{color}

\begin{document}

\title{Imaging disorder-induced scattering centers in quantum Hall incompressible strip}
\author{Y.H. Wang}
\email{yhwang.q@gmail.com}
\affiliation{%
Graduate School of Sciences, Tohoku University, Sendai 980-8578, Japan
}%

\author{K. Hashimoto}%
\affiliation{%
Graduate School of Sciences, Tohoku University, Sendai 980-8578, Japan
}
\affiliation{%
Centre for Spintronics Research Network, Tohoku University, Sendai 980-8578, Japan}

\author{T. Tomimatsu}%
\affiliation{%
Graduate School of Sciences, Tohoku University, Sendai 980-8578, Japan
}

\author{Y. Hirayama}%
\affiliation{%
Graduate School of Sciences, Tohoku University, Sendai 980-8578, Japan
}%
\affiliation{%
Centre for Spintronics Research Network, Tohoku University, Sendai 980-8578, Japan}
\affiliation{%
Center for Science and Innovation in Spintronics (Core Research Cluster), Tohoku University, Sendai 980-8577, Japan}

\date{\today}

\begin{abstract}
While the disorder-induced quantum Hall (QH) effect has been studied previously, 
the effect of disorder potential on microscopic features of the integer QH effect remains unclear, 
particularly for the incompressible (IC) strip.
In this research, 
a scanning gate microscope incorporated with the nonequilibrium transport technique 
is used to image the region of QH IC strip that emerges near the sample edge.
It was found that different mobility samples with varying disorder potentials showed 
the same spatial dependence of the IC strip on the filling factor ($\nu$).
In the low-mobility sample alone,
scattering centers such, bright, dark and annular patterns, alternately appear within the IC strip.
These observed patterns are ascribed to inter-LL scattering assisted by resonance tunneling through an impurity bound state.
It is concluded that disorder-induced scattering can be effectively detected using the applied technique in a low-mobility sample.

\end{abstract}

\pacs{Valid PACS appear here}
\maketitle


The quantum Hall (QH) effect, characterized by nondissipative chiral transport,
is robust to potential disorder, similar to other topological effects such as
the quantum spin Hall \cite{QSH2013,QSHEexperiment2018} and anomalous spin Hall effect \cite{XingQAHE2018}.
The key ingredient of the robustness of QH is the incompressible (IC) phase that protects
the counter propagating metallic compressible (C) edge channels from backscattering
in between the channels at the integer filling factor $\nu = i$ ($i$ is the integer)
\cite{CWJ1990,Chang1990,CSG1992,chakraborty2013QHtextbook}.
The bulk IC phase persists owing to potential disorder at $\nu$, which deviates slightly from $\nu = i$.
This disorder-induced QH state has been microscopically confirmed through scanning probe measurements
\cite{Hashi2008QHT,ilani2004SETprobe,Ashioori2005Capacitance}
that showed compressible paddles interspersed with the IC bulk phase, namely, localized states.
With increasing $\nu$, the IC phase shrinks and moves to the edge of the 2DEG \cite{CSG1992},
eventually forming edge IC strips for integer
\cite{suddards2012capacitance,lai2011MicrowaveImpedance,AhlswedeWeis2001Hallpotential,Weis2011,Yacoby1999image}
and fractional \cite{paradiso2012QPCprl} regimes.
In addition to the bulk localized state,
the IC strip may also help protect against backscattering between the C edge channels through the bulk C region.
Hence, nondissipative transport \cite{SiddikiGerhardts2004},
characterized by a longitudinal resistance that vanishes and a quantized Hall conductance,
is widely persistent around $\nu$.

While some amount of disorder can be stabilizing \cite{stabilizeQHE2003}, further strong disorder might significantly disturb the QH state.
The localized states in the QH effect may become delocalized owing to disorder scattering \cite{Sheng_disorder_destroy_QHE},
which can eventually lead to backscattering between the chiral edge channels.
Sufficiently strong disorder can cause the collapse of quantized plateaus
\cite{Sheng_disorder_destroy_QHE,Disorder&FQHE1986,Disorder&FQHE1984}
as well as topical issues, such as the suppression of the quantized conductance of the quantum spin Hall effect \cite{TopoInsulator2Kane2005}.
A powerful tool used to probe local information of backscattering is a scanning gate microscope (SGM),
which can pinpoint the ``hot spot" where the global transport is susceptible to the gating potential \cite{HotSpot2006}.
SGM can be used to observe the inter-plateau region, for example for half filling factor visualized individual scattering centers,
induced by the disorder in the bulk region \cite{GhrapheneSGM,SGMIhnQHE} and on the edge \cite{SGMwoodside}.
To capture backscattering close to the deep QH region ($\nu \sim i$), namely the plateau region,
a conventional SGM requires the electrostatic influence of a large tip voltage $V_{\rm{tip}} \sim {\pm}1$ V
\cite{paradiso2012QPCprl,pascherIhn2014QPC,SGMQPC2005},
moving the IC strip across a narrow channel and hence losing local information due to potential disorder.
Recently, a scanning gate technique, incorporating nonequilibrium transport \cite{Tomimatsuedgestrip}
with minimization of tip influence, probed the backscattering in a high-mobility ($\mu_{\rm{e}} = 130 \ \rm{m^{2}V^{-1}s^{-1}}$) quantum well sample,
visualizing a line pattern indicating homogeneous backscattering across the IC edge strip.

In this study, we focus on the influence of potential disorder on the IC strips at $\nu \sim i$.
To efficiently access disorder-induced scattering,
the nonequilibrium transport assisted scanning gate technique is applied to a low-mobility sample.
We observed the contrast-modulated line pattern along the sample edge was different from the relatively homogeneous contrast of the IC strip observed in the high-mobility sample.
The $\nu$-dependence of the contrast-modulated patterns were found to be consistent with the IC strip pattern,
demonstrating that sharper disorder potential can give a scattering center in the IC region at $\nu \sim i$.

\begin{figure}
\includegraphics[width=1.0\linewidth]{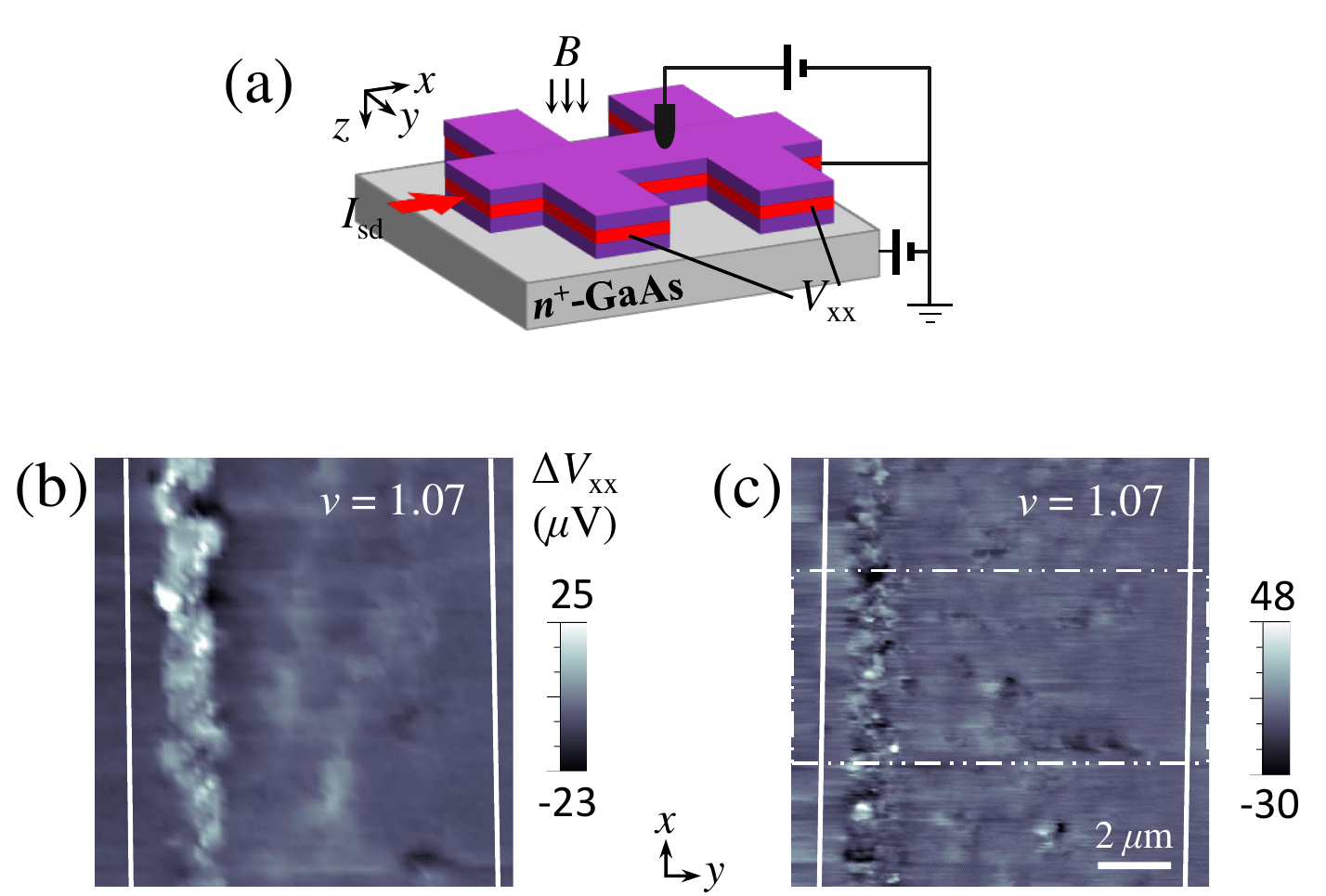}
\caption{(color online) (a) Set up of the SGM. 
$V_{\rm{xx}}$ is recorded by scanning the tip at $V_{\rm{tip}} = 0.2$ V.
$I_{\rm{sd}}$ and $B$ are applied in the $x$ and $z$ directions, respectively. 
A DC bias was applied to the $n^{+}$-GaAs substrate to tune the electron density in 2DEG.
(b) and (c) SGM images of tip-induced $\Delta V_{\rm{xx}}$ 
obtained for (b) the high-mobility sample at $I_{\rm{sd}}= 0.78\ \mu$A
and (c) low-mobility samples at $I_{\rm{sd}}= 0.82\ \mu$A,
at the magnetic field $B = 8$ T.
The white solid lines denote the Hall bar mesa edges.
The line filter was used to remove line noise.
The white dashed box in (c) demarcates the scan window used in Fig. 2.
}
\label{SGM}
\end{figure}

The samples used were 10 $\mu$m width Hall bars, fabricated using 20 nm
GaAs/AlGaAs quantum wells. High- and low-mobility samples were chosen with mobility ($\mu_{\rm{e}}$) of
$130 \ \rm{m^{2}V^{-1}s^{-1}}$ and $25.6 \ \rm{m^{2}V^{-1}s^{-1}}$, respectively,
at an electron density of $n_{\rm{e}} = 1.8\times10^{15}\rm\ m^{-2}$, to ensure different potential disorder.
Figure 1(a) depicts the set-up of our scanning gate method.
The metallic tip, mounted on an atomic force microscope,
was positioned at a constant distance from the surface.
We minimized the tip-induced global depletion (or accumulation) 
by setting $V_{\rm{tip}}$, typically to 0.2 V,
to compensate for the potential mismatch between the tip potential
and global potential of the sample surface.
The current was tuned up to approach the critical point
of the breakdown of the regime to drive the inter-LL tunneling through the innermost IC strip, 
inducing electron tunneling from the edge to the bulk.
This current-induced nonequilibrium transport technique induced backscattering,
while minimizing the influence of current heating \cite{disorde-assistedBreakdownGuven2002}.
Meanwhile, the imposed excess Hall voltage deviated the local chemical potential 
from the ground level of the two-dimensional electron gas (2DEG).
Hence, the tip (a nanoscale top gate) could perturb the microscopic landscape of the potential,
resulting in a modification of the inter-LL tunneling.
In the case of a high-mobility sample,
the nanoscale top gate reduces the width of the IC region by bending the LLs locally and 
enhancing the inter-LL tunneling, particularly through the innermost IC \cite{Tomimatsuedgestrip},
leading to further enhancement of the longitudinal resistance.
The tip-induced resistance change was measured as the longitudinal voltage ($V_{\rm{xx}}$)
at the constant source-drain current ($I_{\rm{sd}}$) using a DC amplifier during the scanning.
We used this scanning gate microscopy technique in particular for a larger disorder system, i.e.,
the low-mobility sample, to pinpoint the scattering sites,
which previously visualized the robust QH IC region \cite{Tomimatsuedgestrip}
in the nonequilibrium condition.
By using a dilution refrigerator and a superconducting magnet, measurements were performed  
at temperatures of 100--300 mK, while varying $\nu$ near $\nu = 1$
by tuning the electron density of the 2DEG via back gating
at a constant magnetic field $B$ = 8 T. 
The experimental results obtained for high- and low-mobility samples were then compared.

Figures 1(b) and 1(c) show the SGM images of $\Delta V_{\rm{xx}}$,
plotted after subtracting the background.
The current flows from down to up,
and the magnetic field is applied perpendicular to the 2DEG plane.
For the high-mobility sample (Fig. 1(b)) at the higher chemical potential side,
a bright line strip can be observed along and near the left Hall-bar edge,
which is attributed to the IC strip \cite{Tomimatsuedgestrip}.
For the low-mobility sample (Fig. 1(c)),
a similar pattern appears along and near the left Hall-bar edge,
although it comprises of alternating bright and
dark spots which differ significantly from the monotonic contrast
in the pattern of the high-mobility sample.

\begin{figure*}[htbp]
\centering
\includegraphics[width=1.0\linewidth]{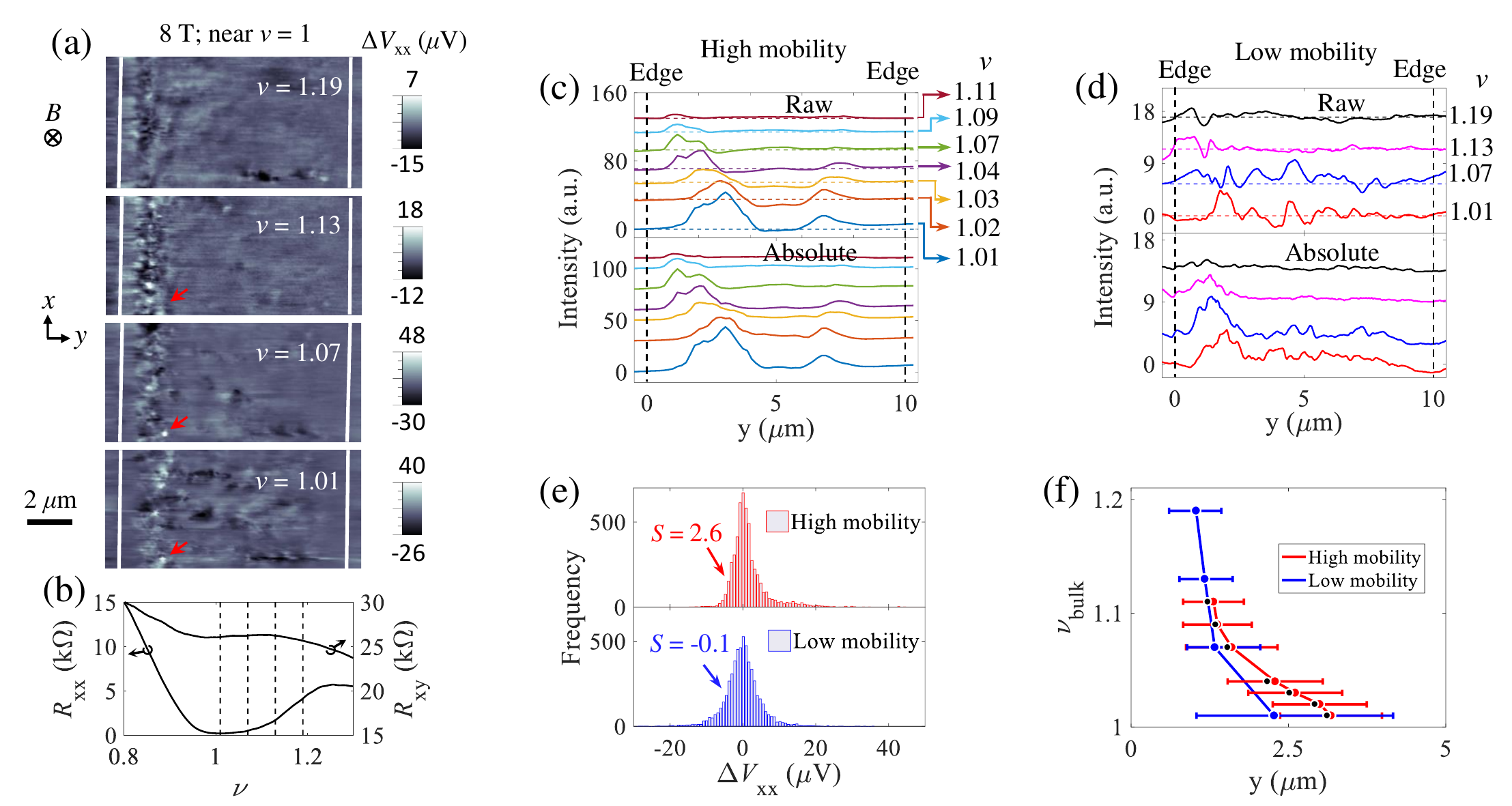}
\caption{(color online)
(a) Representative SGM images at different $\nu$,
at $B = 8$ T near $\nu=1$ for the low-mobility sample.
The scale bar is 2 $\mu$m; the white lines denote the Hall bar mesa edges.
The line filter was used to remove the line noise.
The contrast color scale is obtained by subtracting constant background signal.
(b) QH transport traces for $R_{\rm{xx}}$ and $R_{\rm{xy}}$ expressed as a function of $\nu$ under an equilibrium condition.
The values marked by the dashed lines indicate the $\nu$ chosen in the SGM.
(c) and (d) average line profiles extracted from the high- 
and low-mobility sample images, respectively.
The upper and bottom panels show the average and absolute line profiles, respectively, from the raw data.
The dashed lines indicate the zero lines for each line profile.
(e) Histograms for the high- and low-mobility samples at $\nu=1.07$.
The head of the histogram peak indicates the background in the SGM image,
which has the largest number. The lower and higher intensity sides correspond to the dark and bright spots, respectively.
The asymmetry of the histogram can be measured by skewness, $S$ \cite{wsxm}.
The larger the $S$ value, the more asymmetric the histogram peak.
(f) The position and width are determined
by the distance from the Hall bar edge $y$
and the full width at half maximum $W_{\rm{FWHM}}$, respectively,
which are obtained from absolute line profiles.
The black dots are obtained from the unprocessed line profile for the high-mobility sample.
}
\label{Images}
\end{figure*}

We measured the SGM images at different $\nu$ values around the $R_{\rm{xy}}$ plateau region of $\nu =1$
(marked by the black dashed lines in transport data Fig. 2(b) taken in an equilibrium condition).
The resultant images for the low-mobility sample (Fig. 2(a)) show a line pattern
that moves from the edge to the interior of the Hall bar,
with a reduction of the filling factor toward $\nu =1$.
To compare this position of the line pattern with that for the high-mobility sample,
we took the line profile across the Hall bar
and then spatially averaged the line profile within the measurement area
shown in Fig. 2(a) for the low-mobility sample and a similar size area
for the high-mobility sample (data used in our previous work \cite{Tomimatsuedgestrip}).
The resulting line profiles for the high-mobility sample (the upper panel of Fig. 2(c))
show a peak that moves to the interior of the Hall bar as the $\nu$ value
reduces towards the exact integer $\nu$. This was identified in our previous work \cite{Tomimatsuedgestrip}
as the innermost IC region dependent on $\nu$.
In contrast, the line profiles for the low-mobility sample (the upper panel of Fig. 2(d)) show,
instead of such a clear peak structure,
an oscillatory feature which strengthens near the higher chemical potential side edge.
We speculate that spatial averaging partially canceled out the original oscillatory patterns,
derived from the positive and negative peaks that originate from the bright
and dark spots observed in the 2D image (Fig. 2(a)).

To clarify this point, we performed a statics analysis on the distribution of
the bright and dark spots in the SGM image.
Figure 2(e) shows the intensity histograms of the SGM images for the high-
and low-mobility samples at $\nu = 1.07$.
The skewness ($S$) value \cite{wsxm} for the high-mobility sample
is $S$ = 2.6 (positive asymmetry), indicating that the bright region dominates in the SGM image.
In contrast, for the low-mobility sample, $S$ = -0.1 (almost symmetric distribution),
indicating that the dark and bright spot regions are almost equivalent.
These results lead to the conclusion that the positive and
negative peak structures observed in the low-mobility sample were canceled out 
after spatial averaging and are therefore faint.
To determine the point of the highest intensity of both the positive and negative peaks,
the absolute value of $\Delta V_{\rm{xx}}$ in the line profile was taken, 
followed by spatial averaging over $x$-direction (see the supplementary materials for the detailed method).
The resultant line profiles for the high- and low-mobility samples
are shown in the bottom panels of Fig. 2(c) and 2(d), respectively.
The absolute line profiles for the high-mobility sample (bottom panel of Fig. 2(c))
show a peak structure and $\nu$-dependence 
that are almost identical to those of the unprocessed line profile (upper panel).
By contrast, the line profiles for the low-mobility sample (Fig. 2(d)) significantly change
when taking the absolute values, such that the peak structure near
the high chemical potential side edge becomes obvious (bottom panel).
For further quantitative comparison of the peak structures in the absolute line profile,
the full width at half maximum ($W_{\rm{FWHM}}$) and the position of the peak structures were estimated (see the supplementary materials for the detailed method),
plotted as horizontal bars and dots representing the bulk $\nu$
and position ($y$) (Fig. 2(f)), respectively.
The regions of the peak structures for the low- and high-mobility samples are almost overlapped,
which verifies that the $\nu$-dependence peak structure in the absolute line profile 
for the low-mobility sample exhibits the region of the innermost IC strip.

\begin{figure}
\includegraphics[width=1.0\linewidth]{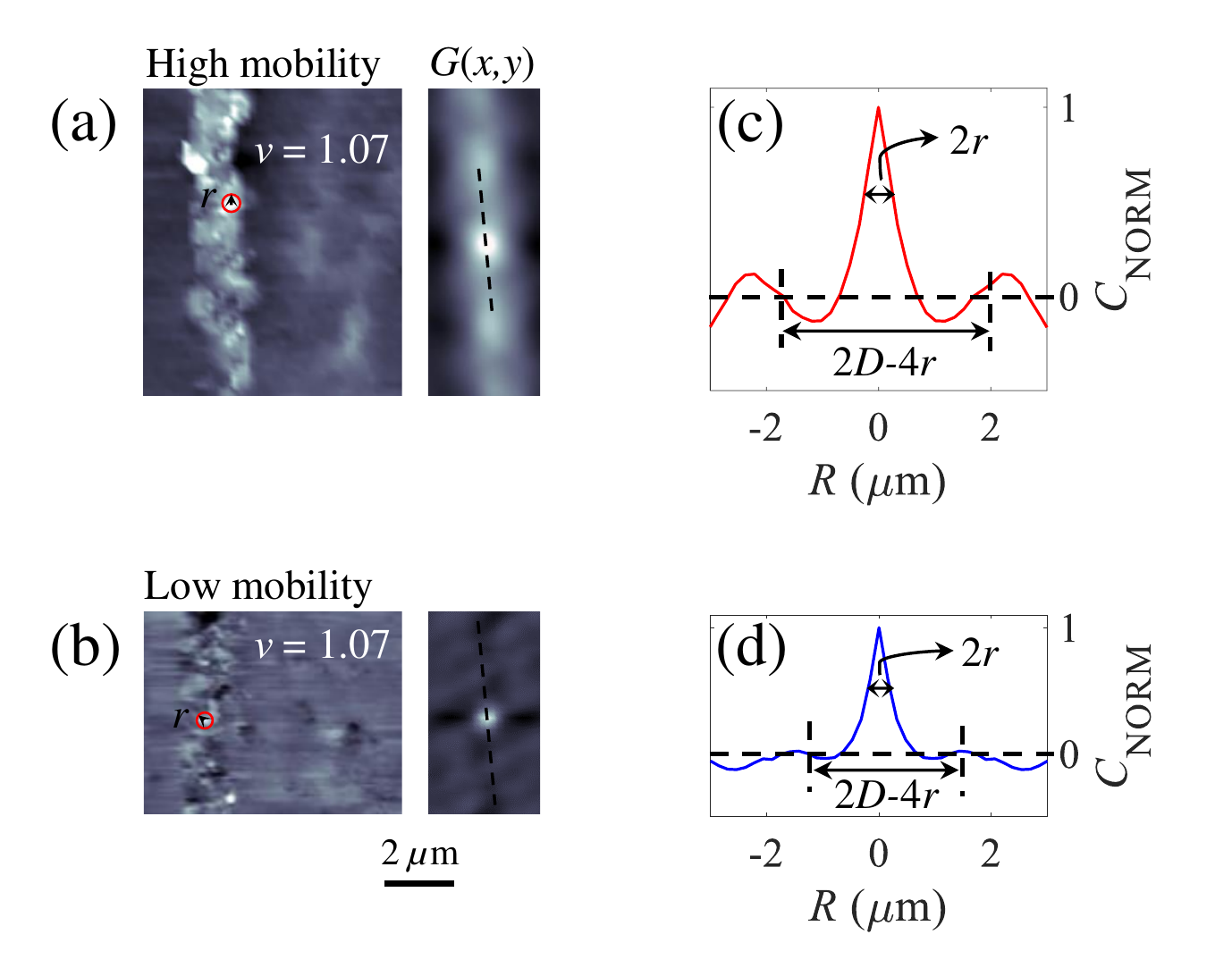}
\caption{(color online) Auto-correlation function obtained for different mobility samples.
(a) and (b) real-space SGM images taken at $\nu = 1.07$ for high- and low-mobility samples, respectively.
Right insets of (a) and (b) represent the $G(x,y)$ of the SGM images.
(c) Line profile of the $G(x,y)$ section for the high-mobility sample, 
which is taken from the peak of the $G(x,y)$.
To avoid influence of the line pattern, 
the cross section was taken in the vertical direction along the line pattern.
(d) Line profile through the $G(x,y)$ peak for the low-mobility sample.}
\label{Auto-correlation}
\end{figure}

Closer examination of the SGM pattern, as marked by circles in Fig. 3(a) and 3(b), 
shows that the size of the bright spot for the low-mobility sample is smaller than that for the high-mobility sample.
To analyze this trend quantitatively,
the two-dimensional auto-correlation function ($G(x,y)$) \cite{wsxm} of the SGM image was calculated.
The right insets of Fig. 3(a) and 3(b) show the representative $G(x,y)$
values which were calculated for the SGM images taken at the $\nu = 1.07$.
The resulting $G(x,y)$ exhibits a disk pattern
which reflects the strong periodic patterns, such as the bright spots, observed in the real-space images 
(marked by the red circle in Fig. 3(a) and (b)). 
Note that a line pattern overlapping the disk pattern for the high-mobility sample
(inset of Fig. 3(a)) is caused by the real-space line pattern overlapping the bright spots.
To compare the characteristics of the real-space hot spots for the high- and low-mobility samples, 
the $G(x,y)$ section was extracted across the disk pattern along the dashed line.
Figure 3(c) and 3(d) show the obtained main peak 
around $R = 0$ with the smaller satellite peaks.
The averaged radius ($r$) of the bright spot and averaged spacing ($D$) between the strong periodic spots were estimated from
the peak width at half maximum of the main peak and the separation between satellite peaks (Fig. 3(c) and 3(d)), 
respectively defined by $2r$ and $2D-4r$ \cite{Autocorrelation,GhrapheneSGM}.
The estimated values are $r = 441$ nm and $D = 2.77$ $\mu$m for the high-mobility sample
and $r = 386$ nm and $D = 1.97$ $\mu$m for the low-mobility sample.
These results indicate that the fine pattern in the IC strip for the 
low-mobility sample is 12 $\%$ smaller and 28 $\%$ denser than that for the high-mobility sample.

\begin{figure}
\includegraphics[width=1.0\linewidth]{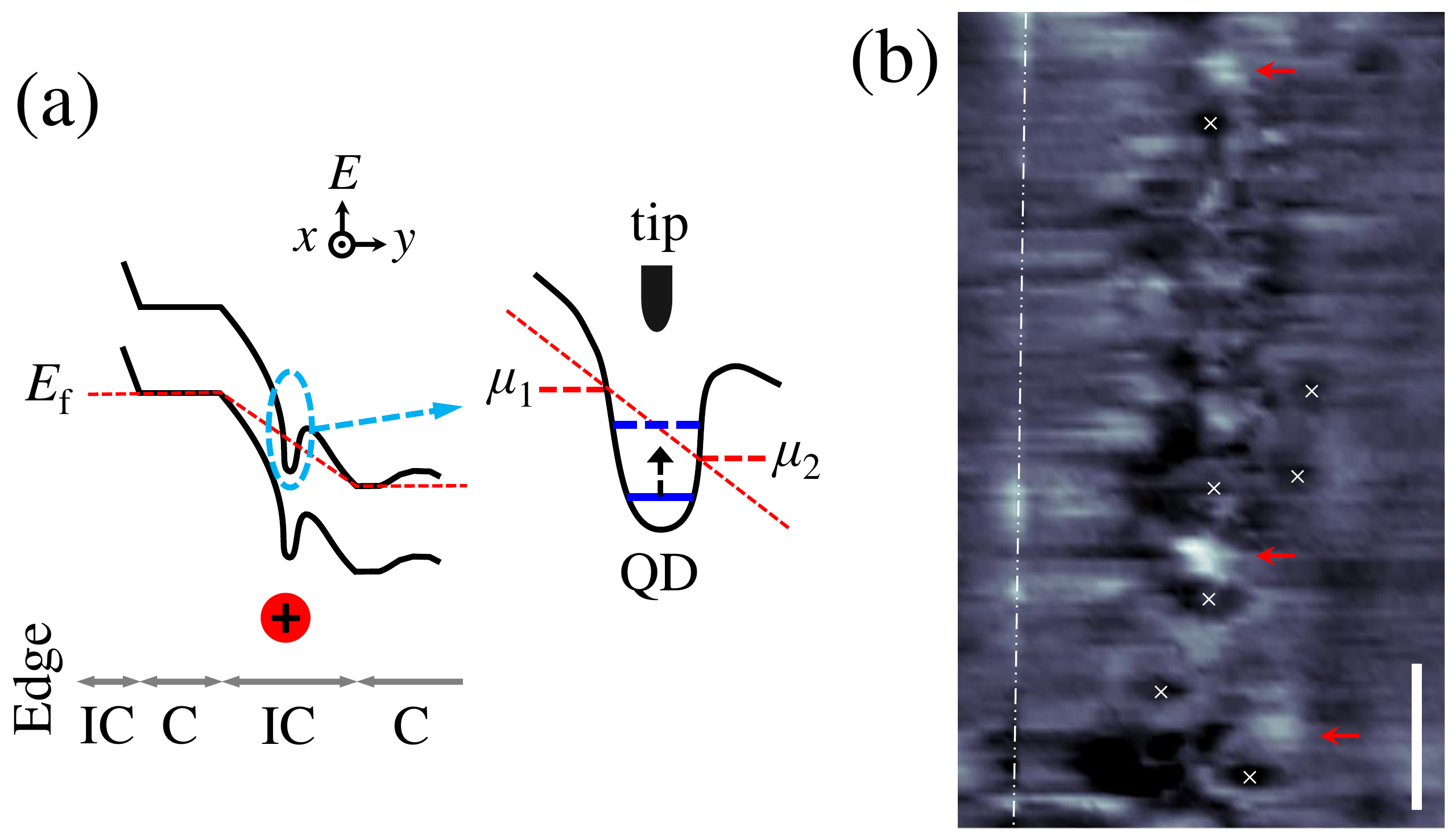}
\caption{(color online)
(a) The schematics of inter-LL with the quantum dot state near the edge for the low-mobility sample.
Right inset is the schematic of QD states for two different levels.
The solid blue line indicates that the initially QD state resides outside the $\Delta\mu$ region.
The blue dashed line indicates the QD state brought into $\Delta \mu$ by the tip gating.
(b) The annular patterns centered around the point (identified by white crosses) located between the pronounced bright spot patterns marked by the red arrows.
The image was taken at $\nu = 1.13$ and includes a part of Fig. 2(a).
The dashed white longitudinal line indicates the sample mesa edge.
Scale bar is 1 $\mu$m.}
\label{spot}
\end{figure}

The obtained fine structures can be interpreted as the pronounced inter-LL scattering centers induced by the disorder potential,
which can have a sharper potential gradient and be denser in the low-mobility sample.
Such potential disorder is primarily induced by charged impurities distributed inhomogeneously 
inside the low-mobility sample.
The disorder-induced potential gradient is significant in the IC region, 
while screened in the compressible region.
For instance, in Fig. 2(a), the bright spot marked by the red arrow appears at the same position 
for $\nu$ = 1.07 and $\nu$ = 1.01 only when the IC strip overlaps with the corresponding position, 
but it remains invisible at $\nu$ = 1.13.

A similar spot pattern was previously observed by Woodside et al. \cite{SGMwoodside}
and interpreted as local inter-LL scattering
caused by the resonance tunneling through a bound state,
which is formed by an impurity-induced potential \cite{Jain1988,SGMwoodside} within the IC strip.
For example, the bound state induced by a positive charged impurity,
in other words a quantum dot (QD) state, resides slightly lower than the chemical potential energy window
($\Delta\mu = \mu_{\rm{1}}-\mu_{\rm{2}}$), as shown in Fig. 4(a).
In our measurement, the small effective potential mismatch between the tip and sample can bring the state into $\Delta\mu$,
enhancing the inter-LL scattering, and hence producing the bright spot.
In contrast, if the QD state that initially locates within $\Delta\mu$ can be detuned by the tip gating,
it can result in the suppression of inter-LL scattering, giving the dark spot observed in the SGM image.
An inhomogeneous distribution of the impurities normal to the 2DEG induces a QD with a different strength confinement.
The resulting random distribution of the QD state in the same energy range in and near $\Delta\mu$
accounts for the bright and dark patterns observed, with almost the same probability.
This imaging mechanism is supported by the annular patterns surrounding the centers marked by crosses in Fig. 4(b),
which can be observed as having a relatively weaker contrast than the pronounced spot patterns (marked by arrows) discussed so far.
A similar pattern was also previously observed and interpreted based on the same origin as the bright/dark spot pattern,
because the bright/dark spot pattern evolves into an annular pattern at high $V_{\rm{tip}}$ \cite{SGMwoodside}.
The high $V_{\rm{tip}}$ was considered to push the QD state into $\Delta\mu$ and hence induce resonance tunneling as the tip approaches the QD center, 
and eventually pushes the state to the outside of $\Delta\mu$ when reaching the QD center,
resulting in annular contrast.
In our disordered sample, the annular contrast can be induced at a low $V_{\rm{tip}}$ by a strong impurity potential that locally increases the effective potential mismatch between the tip and sample.
The average distance between the scattering centers, taking into account both the spot and
annular patterns, is less than half of the average distance of the pronounced periodic patterns,
estimated from $G(x,y)$ ($D$ = 1.97 $\mu$m),
so that the resultant scattering distance is less than the mean free path estimated from the electron density and mobility,
$l = h\mu_{\rm{e}}/e \sqrt{n_{\rm{e}}/2\pi} = 1.79\ \mu m$.
This trend agrees with an analytical result of spin flip scattering between spin-resolved edge channels \cite{Scatterers1999},
which is expected to be larger than the impurity-induced scattering distance observed at $B = 0$ T \cite{SGM_QPC_disorder}
due to suppression of scattering in a high magnetic field \cite{Suppressscattering}.
Such a scenario is likely to occur for the low-mobility sample in which
the charged impurities located in and near the 2DEG layer play an important role.

In contrast, although there are smaller amounts of impurities in and near the quantum well,
the dominant source of disorder potential in the high-mobility sample
can be located relatively far from the 2DEG layer,
for instance, in the barrier layers.
These types of impurities may induce a potential gradient that is relatively smaller than
that induced by the excess Hall voltage imposed by nonequilibrium transport.
When this monotonic potential is disturbed by the small tip potential,
the total tunneling current is strongly enhanced owing to its nonlinear characteristics \cite{Tomimatsuedgestrip}.
Therefore, the bright IC strip is pivotal for the high-mobility sample.

In conclusion, nonequilibrium transport assisted scanning gate microscopy was successfully
applied to detect an IC strip in a low-mobility sample as well as high-mobility sample.
The disorder potential strongly modulates the QH IC phase, 
providing significantly different microscopic perspectives for the different mobility samples.
This work can be extended to study the microscopic influences of potential disorder
on fractional quantum Hall IC strips and IC domain structures \cite{Hayakawa2013}.


We thank K. Muraki and NTT for supplying high-quality wafers, 
K. Sato and K. Nagase for the sample preparation.
Y.H.W thanks the (MEXT) scholarship from Japanese government.
K.H. and T.T. acknowledge the JSPS for financial support: 
KAKENHI 17H02728 and 18K04874, respectively. 
Y.H. acknowledges support from the JSPS 
(KAKENHI 15H0587, 15K217270, and 18H01811). 
K.H. and Y.H. thank Tohoku University's GP-Spin program for support.

\bibliographystyle{apsrev4-1}
\input{references.bbl} 
\end{document}

%% file: references.bbl
%